\documentclass[12pt]{article}
\usepackage{subfloat}
\pdfoutput = 1
\usepackage{graphics}
\usepackage{graphicx} 

\usepackage{hyperref}
\hypersetup{colorlinks=true, urlcolor=blue, citecolor=blue}
\usepackage{amsmath}
\usepackage{amsfonts}   
\usepackage{amssymb}
\usepackage{multirow}
\usepackage{hhline}
\usepackage{color}
\begin{document}
\date{Today}
\title{{\bf{\Large An elementary proof of symmetrization postulate in quantum mechanics for a system of particles }}}
\author{ {\bf {\normalsize Diganta Parai}$^{a}$
\thanks{digantaparai93@gmail.com}},\,
{\bf {\normalsize Nikhilesh Maity}$^{b}
$\thanks{nikhileshm@usf.edu.com, nikhileshmaity28@gmail.com}}\
\\
$^{a}$ {\normalsize Tamluk, West Bengal-721636, India}\\[0.2cm] 
$^{b}$ {\normalsize  Department of Physics } \\{\normalsize University  South Florida, Tampa, Florida 33620, USA }\\[0.2cm]
}
\date{}

\maketitle
\begin{abstract}
\noindent According to symmetrization postulate for a system of identical particles, wave function has to be  completely symmetric or completely anti-symmetric. In this paper we want to mathematically justify this postulate ignoring the spin part of wave function in three dimension. For a system of N identical particles, if the solution to the governing Schrodinger equation meets these criteria: a) the probability density remains invariant when any two particle positions are exchanged over time, b) the wave function is continuous and has a continuous gradient, and the system exhibits the following characteristics: c) the configuration space, which is 3N dimensional, is connected, and d) the potential term in the Hamiltonian is invariant under the exchange of any two particle positions, then the wave function must be either totally symmetric or totally antisymmetric over time.
\end{abstract}
\vskip 1cm
\section{Introduction}

\noindent One of the fundamental postulates in quantum mechanics is the symmetrization postulate. This states that physically realizable wave functions of identical particles must
be either symmetric (Bose-Einstein) or antisymmetric (Fermi-Dirac). Their are many complicated mathematical proofs\cite{1}-\cite{4} and motivational approaches\cite{5}-\cite{7} appeared in the literature about this postulate.  Here, we shall attempt to prove this postulate in an elementary manner, allowing graduate students to easily grasp the mathematical reasoning behind this non-trivial postulate. In this paper we shall consider a certain three dimensional connected finite system and ignore the spin part of the wave function for showing that under the conditions stated above in the abstract, this postulate is nothing but an unavoidable consequence of quantum mechanics for N identical particle scenario. In this proof we do not require the existence of non degenerate energy level or properties involving nodes of wave function( as pointed out in Girardeaus work\cite{3}).  

\noindent  Paper is organized as follows. In section-2, first we shall start with N particle Schrodinger equation then by employing the symmetry of interaction term we shall deduce that under exchange of any two particles coordinate in wave function results $e^{if(t)}$ times non exchanged wave function. Then we will prove that that $f(t)$ is nothing but a constant. After that finally we show that wave function can be only totally anti-symmetric or totally symmetric in nature. In section-3 we extend our analysis for a system of identical particles subjected to electromagnetic field. In section-4 we will conclude.

\section{Proof of the symmetrization postulate }
In most of the books\cite{8}, the symmetry postulate (for two particles) is demonstrated as following way ($\vec{r_1}$ and $\vec{r_2}$ are position of two particles)
\begin{eqnarray}
   \left | \psi(\vec{r_1},\vec{r_2},t)\right |^2= \left | \psi(\vec{r_2},\vec{r_1},t)\right |^2\nonumber\\
   \Rightarrow  \psi(\vec{r_2},\vec{r_1},t)= e^{i\theta}\psi(\vec{r_1},\vec{r_2},t),\nonumber
\end{eqnarray}
 where $\theta$ is assumed to be a constant. Exchange operator P gives
\begin{eqnarray}
    &P&\psi(\vec{r_1},\vec{r_2},t)=\psi(\vec{r_2},\vec{r_1},t)=e^{i\theta}\psi(\vec{r_1},\vec{r_2},t)\nonumber\\
    &\Rightarrow& P^2\psi(\vec{r_1},\vec{r_2},t)=e^{i\theta}P\psi(\vec{r_1},\vec{r_2},t)=e^{2i\theta}\psi(\vec{r_1},\vec{r_2},t)\nonumber\\
    &\Rightarrow& e^{2i\theta}=1\Rightarrow \theta =0,\pi\Rightarrow \psi(\vec{r_2},\vec{r_1},t)=\pm \psi(\vec{r_1},\vec{r_2},t).\nonumber
\end{eqnarray}
So it has been concluded that identical particle wave function can be only two types.\\
\\
But in principle $\theta$ can be a function of $\vec{r_1},\vec{r_2},t$. Then to prove symmetrization postulate along above line of reasoning, we have to show that $\theta(\vec{r_1},\vec{r_2},t)$ is nothing but a constant. In this paper we shall prove that since $\psi$ satisfies Schrodinger equation, $\theta(\vec{r_1},\vec{r_2},t)$ is indeed a constant.
Here we denote $\vec{r_i}$ as position of i th particle. Schrodinger equation of the system is given as
\begin{eqnarray}
    \left[-\frac{\hbar^2}{2m}\sum_{k=1}^{N}\nabla^2_k+V(\vec{r}_1...\vec{r}_{i-1},\vec{r}_i,\vec{r}_{i+1},...\vec{r}_{j-1},\vec{r}_j,\vec{r}_{j+1},...\vec{r}_N,t)\right]\psi(\vec{r}_1...\vec{r}_{i-1},\vec{r}_i,\vec{r}_{i+1},...\vec{r}_{j-1},\vec{r}_j,\vec{r}_{j+1},...\vec{r}_N,t)\nonumber\\
    =i\hbar\frac{\partial}{\partial t}\psi(\vec{r}_1...\vec{r}_{i-1},\vec{r}_i,\vec{r}_{i+1},...\vec{r}_{j-1},\vec{r}_j,\vec{r}_{j+1},...\vec{r}_N,t).\nonumber\\
\end{eqnarray}
From above equation we can write another equation
\begin{eqnarray}
     \left[-\frac{\hbar^2}{2m}\sum_{k=1}^{N}\nabla^2_k+V(\vec{r}_1...\vec{r}_{i-1},\vec{r}_j,\vec{r}_{i+1},...\vec{r}_{j-1},\vec{r}_i,\vec{r}_{j+1},...\vec{r}_N,t)\right]\psi(\vec{r}_1...\vec{r}_{i-1},\vec{r}_j,\vec{r}_{i+1},...\vec{r}_{j-1},\vec{r}_i,\vec{r}_{j+1},...\vec{r}_N,t)\nonumber\\
    =i\hbar\frac{\partial}{\partial t}\psi(\vec{r}_1...\vec{r}_{i-1},\vec{r}_j,\vec{r}_{i+1},...\vec{r}_{j-1},\vec{r}_i,\vec{r}_{j+1},...\vec{r}_N,t).\nonumber\\
\end{eqnarray}
Now according to our assumption (d)
\begin{equation}
  V(\vec{r}_1...\vec{r}_{i-1},\vec{r}_i,\vec{r}_{i+1},...\vec{r}_{j-1},\vec{r}_j,\vec{r}_{j+1},...\vec{r}_N,t)=V(\vec{r}_1...\vec{r}_{i-1},\vec{r}_j,\vec{r}_{i+1},...\vec{r}_{j-1},\vec{r}_i,\vec{r}_{j+1},...\vec{r}_N,t).
\end{equation}
Denote
\begin{eqnarray}
    \psi_{ij}(t)=\psi(\vec{r}_1...\vec{r}_{i-1},\vec{r}_i,\vec{r}_{i+1},...\vec{r}_{j-1},\vec{r}_j,\vec{r}_{j+1},...\vec{r}_N,t),\\
    \psi_{ji}(t)=\psi(\vec{r}_1...\vec{r}_{i-1},\vec{r}_j,\vec{r}_{i+1},...\vec{r}_{j-1},\vec{r}_i,\vec{r}_{j+1},...\vec{r}_N,t).    
\end{eqnarray}
Schrodinger equation is given by
\begin{equation}
    H\psi_{ij}(t)=i\hbar\frac{\partial\psi_{ij}(t)}{\partial t}
    \label{sc0}
\end{equation}
and
\begin{equation}
    H\psi_{ji}(t)=i\hbar\frac{\partial\psi_{ji}(t)}{\partial t}.
    \label{ji}
\end{equation}
Define exchange operator as
\begin{equation}
    \psi_{ji}(t)=P_{ij}\psi_{ij}(t).
\end{equation}
Now equation(\ref{ji}) can be written as
\begin{equation}
    H\left\{P_{ij}\psi_{ij}(t)\right\}=i\hbar\frac{\partial}{\partial t}\left\{P_{ij}\psi_{ij}(t)\right\}.
    \label{schrodinger}
\end{equation}
According to our assumption (a) 
\begin{eqnarray}
    &&\left |\psi_{ji}(t) \right |^2= \left |\psi_{ij}(t) \right |^2\nonumber\\
    &\Rightarrow & \left |P_{ij}\psi_{ij}(t) \right |^2= \left |\psi_{ij}(t) \right |^2\nonumber\\
    &\Rightarrow & P_{ij}\psi_{ij}(t)=e^{i\varphi((\vec{r}_1...\vec{r}_{i-1},\vec{r}_i,\vec{r}_{i+1},...\vec{r}_{j-1},\vec{r}_j,\vec{r}_{j+1},...\vec{r}_N,t)}\psi_{ij}(t)\equiv e^{i\varphi_{ij}(t)}\psi_{ij}(t)\nonumber\\
    &\Rightarrow & P_{ij}^2\psi_{ij}(t)=e^{i\varphi(\vec{r}_1...\vec{r}_{i-1},\vec{r}_j,\vec{r}_{i+1},...\vec{r}_{j-1},\vec{r}_i,\vec{r}_{j+1},...\vec{r}_N,t)}\psi_{ji}(t)\equiv e^{i\varphi_{ji}(t)}\psi_{ji}(t)\nonumber\\
    &\Rightarrow & \psi_{ij}(t)=e^{i(\varphi_{ij}(t)+\varphi_{ji}(t))}\psi_{ij}(t)\nonumber\\
    &\Rightarrow & \varphi_{ji}(t)=-\varphi_{ij}(t)+2n\pi.
\end{eqnarray}
Now we will use equation(\ref{schrodinger}) and
\begin{equation}
    P_{ij}\psi_{ij}(t)=e^{\varphi_{ij}(t)}\psi_{ij},
    \label{exchange}
\end{equation}
to find some consequences.
\begin{eqnarray}
    \vec{\nabla_l}(P_{ij}\psi_{ij}(t))=i e^{i\varphi_{ij}(t)}\psi_{ij}(t)\vec{\nabla_l}\varphi_{ij}(t)+e^{i\varphi_{ij}(t)}\vec{\nabla_l}\psi_{ij}(t)\nonumber\\
    \Rightarrow \nabla_{l}^2(P_{ij}\psi_{ij}(t))=e^{i\varphi_{ij}(t)}\nabla_{l}^2\psi_{ij}(t)+2ie^{i\varphi_{ij}(t)}\vec{\nabla_l}\varphi_{ij}(t)\cdot
     \vec{\nabla_l}\psi_{ij}(t)\nonumber\\
     -e^{i\varphi_{ij}(t)}(\vec{\nabla_l}\varphi_{ij}(t))^2\psi_{ij}(t)+ie^{i\varphi_{ij}(t)}(\nabla_{l}^2\varphi_{ij}(t))\psi_{ij}(t)
     \label{del}
\end{eqnarray}
Now we we explicitly write the equation(\ref{schrodinger}) 
\begin{eqnarray}
    -\frac{\hbar^2}{2m}\left(\sum_{l=1}^{N}\nabla_{l}^2\right)(P_{ij}\psi_{ij}(t))+V(\vec{r_1},...\vec{r_N},t)P_{ij}\psi_{ij}(t)=i\hbar\frac{\partial}{\partial t}\left\{P_{ij}\psi_{ij}(t)\right\}.
    \label{sc1}
\end{eqnarray}
Putting equation(\ref{del}) in equation (\ref{sc1}) we get

\begin{eqnarray}
    &-&\frac{\hbar^2}{2m}\sum_{l=1}^{N}\Big[e^{i\varphi_{ij}(t)}\nabla_{l}^2\psi_{ij}(t)+2ie^{i\varphi_{ij}(t)}\vec{\nabla_l}\varphi_{ij}(t)\cdot\vec{\nabla_l}\psi_{ij}(t)\nonumber\\
    &-&e^{i\varphi_{ij}(t)}(\vec{\nabla_l}\varphi_{ij})^2\psi_{ij}+ie^{i\varphi_{ij}(t)}(\nabla_{l}^2\varphi_{ij}(t))\psi_{ij}(t)\Big]
    +V(\vec{r_1},,,\vec{r_N},t)e^{i\varphi_{ij}(t)}\psi_{ij}(t)\nonumber\\
    &=&i\hbar\left\{ie^{i\varphi_{ij}(t)}\psi_{ij}(t)\frac{\partial \varphi_{ij}(t)}{\partial t}+e^{i\varphi_{ij}(t)}\frac{\partial \psi_{ij}(t)}{\partial t}\right\}.
\end{eqnarray}
Using equation(\ref{sc0}), from above equation we get
\begin{eqnarray}
    \frac{\hbar}{2 m}\sum_{l=1}^{N}\left\{2i(\vec{\nabla_l}\varphi_{ij}(t))\cdot(\vec{\nabla_l}\psi_{ij}(t))-\psi_{ij}(t)(\vec{\nabla_l}\varphi_{ij}(t))^2+i\psi_{ij}(t)\nabla_{l}^2\varphi_{ij}(t)\right\}=\psi_{ij}(t)\frac{\partial \varphi_{ij}(t)}{\partial t}.\nonumber\\
\end{eqnarray}
Multiplying above equation with $\psi_{ij}^*(t)$
\begin{eqnarray}
    \frac{\hbar}{2 m}\sum_{l=1}^{N}\left\{2i(\vec{\nabla_l}\varphi_{ij}(t))\cdot(\psi_{ij}^*(t)\vec{\nabla_l}\psi_{ij}(t))-\left|\psi_{ij}(t)\right|^2(\vec{\nabla_l}\varphi_{ij}(t))^2+i\left|\psi_{ij}(t)\right|^2\nabla_{l}^2\varphi_{ij}(t)\right\}=\left|\psi_{ij}(t)\right|^2\frac{\partial \varphi_{ij}(t)}{\partial t}.\nonumber\\
    \label{+}
\end{eqnarray}
Now taking complex conjugate of above equation(\ref{+})
\begin{eqnarray}
    \frac{\hbar}{2 m}\sum_{l=1}^{N}\left\{-2i(\vec{\nabla_l}\varphi_{ij}(t))\cdot(\psi_{ij}(t)\vec{\nabla_l}\psi_{ij}^*(t))-\left|\psi_{ij}(t)\right|^2(\vec{\nabla_l}\varphi_{ij}(t))^2-i\left|\psi_{ij}(t)\right|^2\nabla_{l}^2\varphi_{ij}(t)\right\}=\left|\psi_{ij}(t)\right|^2\frac{\partial \varphi_{ij}(t)}{\partial t}.\nonumber\\
    \label{-}
\end{eqnarray}
Adding eq.(\ref{+}) and eq.(\ref{-}) we find
\begin{eqnarray}
    &&\frac{\hbar}{2m}\sum_{l=1}^{N}\left[i(\vec{\nabla_l}\varphi_{ij}(t))\cdot\left\{\psi_{ij}^*(t)\vec{\nabla_l}\psi_{ij}(t)-\psi_{ij}(t)\vec{\nabla_l}\psi_{ij}^*(t)\right\}-\left|\psi_{ij}(t)\right|^2(\vec{\nabla_l}\varphi_{ij}(t))^2\right]=\left|\psi_{ij}(t)\right|^2\frac{\partial\varphi_{ij}(t)}{\partial t}\nonumber\\
    &\Rightarrow &\frac{i\hbar}{2m}\sum_{l=1}^{N}\vec{\nabla_l}\cdot\left\{\varphi_{ij}(t)(\psi_{ij}^*(t)\vec{\nabla_l}\psi_{ij}(t)-\psi_{ij}(t)\vec{\nabla_l}\psi_{ij}^*(t))\right\}\nonumber\\
    &-&\frac{i\hbar}{2m}\sum_{l=1}^{N}\left[\varphi_{ij}(t)\vec{\nabla_l}\cdot (\psi_{ij}^*(t)\vec{\nabla_l}\psi_{ij}(t)-\psi_{ij}(t)\vec{\nabla_l}\psi_{ij}^*(t)) \right]
    -\frac{\hbar}{2m}\sum_{l=1}^{N}\left|\psi_{ij}(t)\right|^2(\vec{\nabla_l}\varphi_{ij}(t))^2=\left|\psi_{ij}(t)\right|^2\frac{\partial \varphi_{ij}(t)}{\partial t}.\nonumber\\
    \label{A+B}
\end{eqnarray}
The continuity equation we get
\begin{equation}
    \frac{\partial }{\partial t}(\psi_{ij}^*(t)\psi_{ij}(t))=\frac{i\hbar}{2m}\sum_{l=1}^{N}\vec{\nabla_l}\cdot (\psi_{ij}^*(t)\vec{\nabla_l}\psi_{ij}(t)-\psi_{ij}(t)\vec{\nabla_l}\psi_{ij}^*(t)).
    \label{con}
\end{equation}
Using equation(\ref{con}) in the equation(\ref{A+B}) we get
\begin{eqnarray}
    &&\frac{i\hbar}{2m}\sum_{l=1}^{N}\vec{\nabla_l}\cdot\left\{\varphi_{ij}(t)(\psi_{ij}^*(t)\vec{\nabla_l}\psi_{ij}(t)-\psi_{ij}(t)\vec{\nabla_l}\psi_{ij}^*(t))\right\}\nonumber\\
    &-&\varphi_{ij}(t)\frac{\partial }{\partial t}\left|\psi_{ij}(t)\right|^2
    -\frac{\hbar}{2m}\sum_{l=1}^{N}\left|\psi_{ij}(t)\right|^2(\vec{\nabla_l}\varphi_{ij}(t))^2=\left|\psi_{ij}(t)\right|^2\frac{\partial \varphi_{ij}(t)}{\partial t}\nonumber\\
    &\Rightarrow &\frac{i\hbar}{2m}\sum_{l=1}^{N}\vec{\nabla_l}\cdot\left\{\varphi_{ij}(t)(\psi_{ij}^*(t)\vec{\nabla_l}\psi_{ij}(t)-\psi_{ij}(t)\vec{\nabla_l}\psi_{ij}^*(t))\right\}\nonumber\\
    &=&\frac{\hbar}{2m}\sum_{l=1}^{N}\left|\psi_{ij}(t)\right|^2(\vec{\nabla_l}\varphi_{ij}(t))^2+\frac{\partial}{\partial t}\left[\left|\psi_{ij}(t)\right|^2\varphi_{ij}(t)\right].
\end{eqnarray}
Now we shall integrate above equation throughout the whole configuration space and we get (employing the boundary condition is that over the boundary $\psi$ is zero )
\begin{eqnarray}
    -\frac{\hbar}{2m}\int \left|\psi_{ij}(t)\right|^2\left[\sum_{l=1}^{N}(\vec{\nabla_l}\varphi_{ij}(t))^2\right]\prod_{k=1}^{N}d^3\vec{r_k}&=&\frac{d}{dt}\int \left|\psi_{ij}(t)\right|^2\varphi_{ij}(t)\prod_{k=1}^{N}d^3\vec{r_k}\nonumber\\
    &=& \frac{d}{dt}\int \left|\psi_{ji}(t)\right|^2\varphi_{ji}(t)\prod_{k=1}^{N}d^3\vec{r_k}\nonumber\\
    &=&\frac{d}{dt}\int \left|\psi_{ij}(t)\right|^2(-\varphi_{ij}(t)+2n\pi)\prod_{k=1}^{N}d^3\vec{r_k}\nonumber\\
    &=&-\frac{d}{dt}\int  \left|\psi_{ij}(t)\right|^2\varphi_{ij}(t)\prod_{k=1}^{N}d^3\vec{r_k}.
\end{eqnarray}
So we have conclude from above calculation is that
\begin{equation}
    \int \left|\psi_{ij}(t)\right|^2\left[\sum_{l=1}^{N}(\vec{\nabla_l}\varphi_{ij}(t))^2\right]\prod_{k=1}^{N}d^3\vec{r_k}=0
\end{equation}
So any continuous functions that is function of $\vec{r_1},...,\vec{r_N}$ can not be $\varphi_{ij}(t)$. Situation in this point is
\begin{equation}
    P_{ij}\psi_{ij}(t)=e^{i\varphi(t)}\psi_{ij}(t)=\psi_{ji}(t)
    \label{tform}
\end{equation}
Let define
\begin{eqnarray}
    S(t)&=&\int\psi^*_{ij}(t)\psi_{ji}(t)\prod_{k=1}^{N}d^3\vec{r_k}.\nonumber\\
    \frac{d S}{dt}&=&\int\left[\frac{\partial \psi_{ij}^*(t)}{\partial t}\psi_{ji}(t)+\psi_{ij}^*(t)\frac{\partial \psi_{ji}}{\partial t}\right]\prod_{k=1}^{N}d^3\vec{r_k}\nonumber\\
    &=&\frac{i}{\hbar}\int\left[\psi_{ji}(t)H\psi_{ij}^*(t)-\psi_{ij}^*(t)H\psi_{ji}(t)\right]\prod_{k=1}^{N}d^3\vec{r_k}\nonumber
\end{eqnarray}    
\begin{eqnarray}
   \frac{d S}{dt} &=&-\frac{i\hbar}{2m}\int\sum_{l=1}^{N}\left[\psi_{ji}(t)\nabla_l^2\psi_{ij}^*(t)-\psi_{ij}^*(t)\nabla_l^2\psi_{ji}(t)\right]\prod_{k=1}^{N}d^3\vec{r_k}\nonumber\\
    &=&-\frac{i\hbar}{2m}\sum_{l=1}^{N}\left[\int \vec{\nabla_l}\cdot\left\{\psi_{ji}(t)\vec{\nabla_l}\psi_{ij}^*(t)-\psi_{ij}^*(t)\vec{\nabla_l}\psi_{ji}(t)\right\}\prod_{k=1}^{N}d^3\vec{r_k}\right]\nonumber\\
    &=&-\frac{i\hbar}{2m}\sum_{l=1}^{N}\left[\int \vec{\nabla_l}\cdot\left\{\psi_{ji}(t)\vec{\nabla_l}\psi_{ij}^*(t)\right\}\prod_{k=1}^{N}d^3\vec{r_k}\right]\nonumber\\
    &=&-\frac{i\hbar}{2m}\sum_{l=1}^{N}\left[\int \vec{\nabla_l}\cdot\left\{\psi_{ij}(t)\vec{\nabla_l}\psi_{ji}^*(t)\right\}\prod_{k=1}^{N}d^3\vec{r_k}\right]=0
    \label{S}
\end{eqnarray}
 The above calculation implies $S(t)$ is constant. Using equation(\ref{tform}) we get
 \begin{eqnarray}
   &&\frac{d}{dt}\int\psi^*_{ij}(t)\psi_{ji}(t)\prod_{k=1}^{N}d^3\vec{r_k}=0\nonumber\\
   &\Rightarrow& \frac{d}{dt}\int \psi_{ij}^*(t)e^{i\varphi(t)}\psi_{ij}(t)\prod_{k=1}^{N}d^3\vec{r_k}=0\nonumber\\
   &\Rightarrow& \frac{d}{dt} e^{i\varphi(t)}=0\nonumber\\
   &\Rightarrow& \varphi(t)=const\equiv A.
 \end{eqnarray}
 So,
 \begin{eqnarray}
     &&P_{ij}\psi_{ij}(t)=e^{i A}\psi_{ij}(t)\nonumber\\
  &\Rightarrow&  P_{ij}^2\psi_{ij}(t)=e^{i A}\psi_{ji}(t)=e^{2iA}\psi_{ij}(t)\nonumber\\
  &\Rightarrow& e^{2iA}=1=e^{2\pi i n}\Rightarrow A=n\pi\nonumber\\
  &\Rightarrow& P_{ij}\psi_{ij}(t)=e^{in\pi}\psi_{ij}(t)\nonumber\\
  &\Rightarrow& \psi_{ji}(t)=\pm\psi_{ij}(t).
  \end{eqnarray}
  Now we want to show that if at any time $t=t_0$, $\psi_{ji}(t_0)=-\psi_{ij}(t_0)$ it fixes that $\psi_{ji}(t)=-\psi_{ij}(t)$ and same conclusion applied for symmetric case also.
  \begin{eqnarray}
      \int\left|\psi_{ij}(t)+\psi_{ji}(t)\right|^2\prod_{k=1}^{N}d^3\vec{r_k}=2+2\int\psi_{ij}^*(t)\psi_{ji}(t)\prod_{k=1}^{N}d^3\vec{r_k}
  \end{eqnarray}
  Now second term in the right hand side is constant with time. It is very simple that at $t=t_{0}$ that term is $-1$. So,
  \begin{eqnarray}
      \int\left|\psi_{ij}(t)+\psi_{ji}(t)\right|^2\prod_{k=1}^{N}d^3\vec{r_k}=0\nonumber\\
      \Rightarrow \psi_{ij}(t)=-\psi_{ji}(t).
  \end{eqnarray}
  Similar line of reasoning can be extended for symmetric case.\\
  
  \noindent So we have proved that under exchange of any two particles position, wave function will be unchanged or sign flip will occur. Now we want to show that if wave function is symmetric under exchange of two particular particles coordinate, wave function has to be symmetric under exchange of any two particles coordinate. For demonstration take a three particle scenario. Wave function will be $\psi(\vec{r_1},\vec{r_2},\vec{r_3},t)$. Our assumption is suppose exchange of first slot and second slot, wave function is symmetric and under exchange of second slot and third slot, wave function is antisymmetric then
  \begin{eqnarray}
&\psi&(\vec{r_1},\vec{r_2},\vec{r_3},t)=\psi(\vec{r_2},\vec{r_1},\vec{r_3},t)=-\psi(\vec{r_2},\vec{r_3},\vec{r_1},t)\nonumber\\
=&-&\psi(\vec{r_3},\vec{r_2},\vec{r_1},t)=\psi(\vec{r_3},\vec{r_1},\vec{r_2},t)=\psi(\vec{r_1},\vec{r_3},\vec{r_2},t)=-\psi(\vec{r_1},\vec{r_2},\vec{r_3},t)\nonumber\\
&\Rightarrow & \psi(\vec{r_1},\vec{r_2},\vec{r_3},t)=0.
  \end{eqnarray}

  So many particle wave function has to be totally symmetric or totally antisymmetric.

\section{Proof of symmetrization postulate for a system of identical particles in electromagnetic field} 

\noindent In the previous discussion, we have investigated the symmetrization postulate for a system of particles governed by the normal Schrodinger equation with a symmetric interaction term. Now in the presence of an electromagnetic field, $\vec{P}$ will change with $\vec{P}-q\vec{A}$ in the Schrodinger equation. Resulting equation can not be identified with normal Schrodinger equation with a modified interaction term. So we need a separate treatment for this case. Here we will consider the system subjected to electromagnetic field. Sequence of proof will go like previous treatment. Here we will work in coulomb gauge that is

\begin{equation}
    \vec{\nabla}\cdot \vec{A}=0
\end{equation}
\noindent The Schrodinger equation for the system will be given by
\begin{equation}
    \left[\sum_{\mu=1}^{N}\left\{\frac{1}{2m}\left(i\hbar\vec{\nabla }_{\mu}+q\vec{A}_{\mu}\right)^2+U(\vec{r}_{\mu},t)\right\}+V(
\vec{r}_1,...,\vec{r}_N,t)\right]\psi_{ij}(t)=i\hbar\frac{\partial \psi_{ij}(t)}{\partial t},
\label{n1}
\end{equation}
\noindent where $\vec{A}_{\mu}=\vec{A}_{\mu}(\vec{r}_{\mu},t)$, $U$ is electric potential and $V$ is the interaction term satisfying assumption $(d)$. We can rewrite eq.(\ref{n1}) as
\begin{equation}
   \left[\sum_{\mu=1}^{N}\left\{-\frac{\hbar^2}{2m}\nabla_{\mu}^2+\frac{i\hbar q }{m}\vec{A}_{\mu}.\vec{\nabla}_{\mu}+\frac{q^2}{2m}\vec{A}_{\mu}^2\right\}+\bar{V}(\vec{r_1},...,\vec{r_N},t)\right]\psi_{ij}(t)=i\hbar\frac{\partial \psi_{ij}(t)}{\partial t},
   \label{n2}
\end{equation}
where in $\bar{V}$ we have absorbed electric potential term and interaction term. Inspecting the form of eq.(\ref{n2}) we can write anather equation as
\begin{equation}
 \left[\sum_{\mu=1}^{N}\left\{-\frac{\hbar^2}{2m}\nabla_{\mu}^2+\frac{i\hbar q }{m}\vec{A}_{\mu}.\vec{\nabla}_{\mu}+\frac{q^2}{2m}\vec{A}_{\mu}^2\right\}+\bar{V}(\vec{r_1},...,\vec{r_N},t)\right]\psi_{ji}(t)=i\hbar\frac{\partial \psi_{ji}(t)}{\partial t}.  
\end{equation}
According to eq.(\ref{exchange}) we can write from above equation as
\begin{equation}
 \left[\sum_{\mu=1}^{N}\left\{-\frac{\hbar^2}{2m}\nabla_{\mu}^2+\frac{i\hbar q }{m}\vec{A}_{\mu}.\vec{\nabla}_{\mu}+\frac{q^2}{2m}\vec{A}_{\mu}^2\right\}+\bar{V}(\vec{r_1},...,\vec{r_N},t)\right]e^{\varphi_{ij}(t)}\psi_{ij}(t)=i\hbar\frac{\partial }{\partial t}\left\{e^{\varphi_{ij}(t)}\psi_{ij}(t)\right\}.
\end{equation}
Simplifying above equation we get
\begin{eqnarray}
    \sum_{\mu=1}^{N} &\Bigg[&\frac{\hbar}{2 m}\left\{2i(\vec{\nabla_{\mu}}\varphi_{ij}(t))\cdot(\psi_{ij}^*(t)\vec{\nabla_{\mu}}\psi_{ij}(t))-\left|\psi_{ij}(t)\right|^2(\vec{\nabla_{\mu}}\varphi_{ij}(t))^2+i\left|\psi_{ij}(t)\right|^2\nabla_{\mu}^2\varphi_{ij}(t)\right\}\nonumber\\
    &+&\frac{q}{m}\left|\psi_{ij}(t)\right|^2\vec{A}_{\mu}\cdot\left(\vec{\nabla}_{\mu}\varphi_{ij}(t)\right)\Bigg]=\left|\psi_{ij}(t)\right|^2\frac{\partial \varphi_{ij}(t)}{\partial t}.
\end{eqnarray}
Taking complex conjugate of above equation and adding with it, we find
\begin{eqnarray}
    &\Rightarrow &\frac{i\hbar}{2m}\sum_{\mu=1}^{N}\vec{\nabla_{\mu}}\cdot\left\{\varphi_{ij}(t)(\psi_{ij}^*(t)\vec{\nabla_{\mu}}\psi_{ij}(t)-\psi_{ij}(t)\vec{\nabla_{\mu}}\psi_{ij}^*(t))\right\}\nonumber\\
    &-&\frac{i\hbar}{2m}\sum_{\mu=1}^{N}\left[\varphi_{ij}(t)\vec{\nabla_{\mu}}\cdot (\psi_{ij}^*(t)\vec{\nabla_{\mu}}\psi_{ij}(t)-\psi_{ij}(t)\vec{\nabla_{\mu}}\psi_{ij}^*(t)) \right]\nonumber\\
    &-&\frac{\hbar}{2m}\sum_{\mu=1}^{N}\left|\psi_{ij}(t)\right|^2(\vec{\nabla_{\mu}}\varphi_{ij}(t))^2
    +\sum_{\mu=1}^{N}\frac{q}{m}\left|\psi_{ij}(t)\right|^2\vec{A}_{\mu}\cdot\left(\vec{\nabla}_{\mu}\varphi_{ij}(t)\right)
    =\left|\psi_{ij}(t)\right|^2\frac{\partial \varphi_{ij}(t)}{\partial t}.\nonumber\\
    \label{main}
\end{eqnarray}
The continuity equation we get
\begin{equation}
    \frac{\partial }{\partial t}(\psi_{ij}^*(t)\psi_{ij}(t))=\frac{i\hbar}{2m}\sum_{\mu=1}^{N}\vec{\nabla_{\mu}}\cdot (\psi_{ij}^*(t)\vec{\nabla_{\mu}}\psi_{ij}(t)-\psi_{ij}(t)\vec{\nabla_{\mu}}\psi_{ij}^*(t))+\sum_{\mu=1}^{N}\frac{q}{m}\vec{A}_{\mu}\cdot\left(\vec{\nabla}_{\mu}\left|\psi_{ij}(t)\right|^2\right).
    \label{conti}
\end{equation}
Using eq.(\ref{main}) and eq.(\ref{conti}) we find
\begin{eqnarray}
    &\Rightarrow &\frac{i\hbar}{2m}\sum_{\mu=1}^{N}\vec{\nabla_{\mu}}\cdot\left\{\varphi_{ij}(t)(\psi_{ij}^*(t)\vec{\nabla_{\mu}}\psi_{ij}(t)-\psi_{ij}(t)\vec{\nabla_{\mu}}\psi_{ij}^*(t))\right\}\nonumber\\
    &=&\frac{\hbar}{2m}\sum_{\mu=1}^{N}\left|\psi_{ij}(t)\right|^2(\vec{\nabla_{\mu}}\varphi_{ij}(t))^2+\frac{\partial}{\partial t}\left[\left|\psi_{ij}(t)\right|^2\varphi_{ij}(t)\right]-\frac{q}{m}\sum_{\mu=1}^{N}\vec{\nabla}_{\mu}\cdot\left\{\varphi_{ij}(t)\left|\psi_{ij}(t)\right|^2\vec{A}_{\mu}\right\}.\nonumber\\
\end{eqnarray}
Now if we integrate above equation throughout whole configuration space we get
\begin{eqnarray}
    -\frac{\hbar}{2m}\int \left|\psi_{ij}(t)\right|^2\left[\sum_{\mu=1}^{N}(\vec{\nabla_{\mu}}\varphi_{ij}(t))^2\right]\prod_{k=1}^{N}d^3\vec{r_k}
    =\frac{d}{dt}\int \left|\psi_{ij}(t)\right|^2\varphi_{ij}(t)\prod_{k=1}^{N}d^3\vec{r_k},  
\end{eqnarray}
which is same equation that we encountered in section-2 and concluded that
$\varphi_{ij}(t)$ is solely a function of $t$ only. So
\begin{equation}
    \psi_{ji}(t)=e^{\varphi(t)}\psi_{ij}(t).\nonumber
\end{equation}
Similar to the section-2 we will define
\begin{eqnarray}
    S(t)&=&\int\psi^*_{ij}(t)\psi_{ji}(t)\prod_{k=1}^{N}d^3\vec{r_k}.\nonumber\\
    \frac{d S}{dt}&=&\int\left[\frac{\partial \psi_{ij}^*(t)}{\partial t}\psi_{ji}(t)+\psi_{ij}^*(t)\frac{\partial \psi_{ji}}{\partial t}\right]\prod_{k=1}^{N}d^3\vec{r_k}\nonumber\\
    &=&\frac{i}{\hbar}\int\left[\psi_{ji}(t)H^*\psi_{ij}^*(t)-\psi_{ij}^*(t)H\psi_{ji}(t)\right]\prod_{k=1}^{N}d^3\vec{r_k}\nonumber\\
    &=&-\frac{i\hbar}{2m}\int\sum_{\mu=1}^{N}\left[\psi_{ji}(t)\nabla_{\mu}^2\psi_{ij}^*(t)-\psi_{ij}^*(t)\nabla_{\mu}^2\psi_{ji}(t)\right]\prod_{k=1}^{N}d^3\vec{r_k}\nonumber\\
    &+&\frac{q}{m}\int\sum_{\mu=1}^{N}\vec{A}_{\mu}\cdot\vec{\nabla}_{\mu}\left\{\psi_{ji}(t)\psi_{ij}^*(t)\right\}\prod_{k=1}^{N}d^3\vec{r_k}\nonumber\\
    &=&-\frac{i\hbar}{2m}\sum_{\mu=1}^{N}\left[\int \vec{\nabla_{\mu}}\cdot\left\{\psi_{ji}(t)\vec{\nabla_{\mu}}\psi_{ij}^*(t)-\psi_{ij}^*(t)\vec{\nabla_{\mu}}\psi_{ji}(t)\right\}\prod_{k=1}^{N}d^3\vec{r_k}\right]\nonumber\\
    &+&\frac{q}{m}\int\sum_{\mu=1}^{N}\vec{\nabla}_{\mu}\cdot\left\{\psi_{ji}(t)\psi_{ij}^*(t)\vec{A}_{\mu}\right\}\prod_{k=1}^{N}d^3\vec{r_k}\nonumber\\
    &=&-\frac{i\hbar}{2m}\sum_{\mu=1}^{N}\left[\int \vec{\nabla_{\mu}}\cdot\left\{\psi_{ji}(t)\vec{\nabla_{\mu}}\psi_{ij}^*(t)\right\}\prod_{k=1}^{N}d^3\vec{r_k}\right]\nonumber\\
    &=&-\frac{i\hbar}{2m}\sum_{\mu=1}^{N}\left[\int \vec{\nabla_{\mu}}\cdot\left\{\psi_{ij}(t)\vec{\nabla_{\mu}}\psi_{ji}^*(t)\right\}\prod_{k=1}^{N}d^3\vec{r_k}\right]=0
\end{eqnarray}
So we conclude that $S(t)$ is a constant. Now from this point, following the argument of section-2(eq.(\ref{S}) onwards) we can conclude that wave function is either totally symmetric or totally antisymmetric. 
\section{Conclusions}
In this paper starting from a N particle Schrodinger equation satisfying four conditions stated in the abstract we have arrived at the symmetrization postulate. We have explicitly shown that wave function for a system of N identical spinless particles having a symmetric interaction term, can be either totally symmetric or totally anti-symmetric in the exchange of space coordinates. Same conclusion achieved for the identical particle system subjected to electromagnetic field. For a non-interacting system of particles we can deduce Paulies exclusion principal and slatter determinant formula easily from totally antisymmetric property of wave function.

\section*{Acknowledgments}
N.M. acknowledge financial support by the U.S. National Science Foundation under grant No. DMR-2219476.


\begin{thebibliography}{99}
\baselineskip=0.6 cm
\bibitem{1} Flicker, Michael, and Harvey S. Leff. "Symmetrization postulate of quantum mechanics." Physical Review 163.5 (1967): 1353.
\bibitem{2} Girardeau, M. D. "Proof of the symmetrization postulate." Journal of Mathematical Physics 10.7 (1969): 1302-1304.
\bibitem{3}Girardeau, M. D. "Permutation symmetry of many-particle wave functions." Physical Review 139.2B (1965): B500.
\bibitem{4} Messiah, Albert ML, and Oscar W. Greenberg. "Symmetrization postulate and its experimental foundation." Physical Review 136.1B (1964): B248.
\bibitem{5}Bigaj, Tomasz. "How to justify the symmetrization postulate in quantum mechanics." Journal for General Philosophy of Science 53.3 (2022): 239-257.
\bibitem{6} Kaplan, Ilya G. "The Pauli exclusion principle and the problems of its experimental verification." Symmetry 12.2 (2020): 320.
\bibitem{7}Hilborn, Robert C., and Candice L. Yuca. "Identical particles in quantum mechanics revisited." The British journal for the philosophy of science 53.3 (2002): 355-389.
\bibitem{8} Phillips, Anthony C. Introduction to quantum mechanics. John Wiley and Sons, 2013.
\end{thebibliography}
\end{document}